\documentclass[fleqn,usenatbib]{mnras}
\pdfoutput=1 
\usepackage{newtxtext,newtxmath}
\usepackage[T1]{fontenc}
\usepackage{ae,aecompl}
\usepackage{hyperref} 
\usepackage{url}

\usepackage{graphicx}
\usepackage{subcaption}
\def\be{\begin{equation}} 
\def\ee{\end{equation}} 
\def\bea{\begin{eqnarray}} 
\def\eea{\end{eqnarray}} 

\usepackage[table,usenames,dvipsnames]{xcolor}

\title[A Neural Network For Cosmic String Detection]{A Convolutional Neural Network For Cosmic String Detection in CMB Temperature Maps}{}
\author[Ciuca, Hern\'andez, Wolman]{
Razvan Ciuca$^{1,2,3}$\thanks{Email: razvan.ciuca@mail.mcgill.ca}
Oscar F. Hern\'andez$^{1,2}$\thanks{Email: oscarh@physics.mcgill.ca} 
and Michael Wolman$^{2,3}$\thanks{Email: michael.s.wolman@gmail.com}\\
$^{1}$Department of Physics, McGill University, 3600 rue University, Montr\'eal, QC, H3A 2T8, Canada \\
$^{2}$Marianopolis College,  4873 Westmount Ave.,Westmount, QC H3Y 1X9, Canada \\
$^{3}$School of Computer Science, McGill University,  3480 rue University, Montr\'eal, QC, H3A 0E9, Canada
}

\date{}  \pubyear{2017}
\begin{document}
\label{firstpage} \pagerange{\pageref{firstpage}--\pageref{lastpage}}
\maketitle

\begin{abstract}
We present in detail the convolutional neural network used in our previous work to detect cosmic strings in cosmic microwave background (CMB) temperature anisotropy maps. By training this neural network on numerically generated CMB temperature maps, with and without cosmic strings, the network can produce prediction maps that locate the position of the cosmic strings and provide a probabilistic estimate of the value of the string tension $G\mu$. Supplying noiseless simulations of CMB maps with arcmin resolution to the network resulted in the accurate determination both of string locations and string tension for sky maps having strings with string tension as low as $G\mu=5\times10^{-9}$, a result from our previous work. In this work we discuss the numerical details of the code that is publicly available online. Furthermore, we show that though we trained the network with a long straight string toy model, the network performs well with realistic Nambu-Goto simulations. 
\end{abstract}


\section{Introduction}
\label{sec:intro}
Cosmic strings are line-like topological defects, remnants of a high-energy phase transition in the very early Universe which can form in a large class of extensions of the Standard Model.  
The gravitational effects of the string are parametrized by its string tension $G\mu$, a dimensionless constant where $G$ is Newton's gravitational constant, and $\mu$ is the energy per unit length of the string. Currently the~\citet{PlanckCollaboration:2014il} provides the best robust limits on the string tension with $G\mu \lesssim 10^{-7}$ at the 95\% confidence level. 
%

In recent years much research has been done to find a more sensitive probe of cosmic strings in cosmic microwave background (CMB) and 21 cm intensity maps.  As long cosmic strings move, they accrete matter into overdense wakes which perturb the CMB radiation and the 21 cm line in particular. Future 21 cm redshift surveys could observe cosmic string wakes through their distinctive shape in redshift space~\citep{Brandenberger:2010hi, Hernandez:2011ima, Hernandez:2012gz, Hernandez:2014cu, daCunha:2016bo}
or through the cross-correlation between CMB and 21-cm radiation from dark ages~\citep{Berndsen:2010ku}.  Edge and shape detection algorithms such as the Canny algorithm~\citep{Canny:et}, wavelets, and curvelets have been proposed and studied as alternatives to the power spectrum in looking for cosmic strings in these maps.  
A string moving between an observer and the surface of last scattering can lead to a step discontinuity in a CMB temperature anisotropy map through the Gott-Kaiser-Stebbins (GKS) effect of long strings~\citep{Gott:1985eg,Kaiser:1984jg}. 
\cite{Amsel:2008it,Stewart:2009fr, 2010IJMPD..19..183D} used the Canny algorithm to look for GKS edges produced by long strings in the CMB temperature maps simulated. They found more short edges in maps with strings which they interpreted as the disruption of long edges by Gaussian noise. However as shown in~\cite{Ciuca:2017jz} these edges do not necessarily correspond to the string locations. References~\cite{Hergt:2017dr, McEwen:2017cg} used curvelet transforms to analyse simulated CMB temperature maps with noise, and most recently~\cite{VafaeiSadr:2018hh} has combined both Canny and curvelets to place a detection limit of $G\mu\sim 10^{-7}$ on maps with realistic noise.  

All of the methods discussed above can be thought of as a statistic on a sky map. These methods involve choices and it remains unclear whether or not a different choice would improve detection. For example, in the Canny algorithm changing the gradient thresholds changes the number of edges found. In the wavelet and curvelet analysis there is a choice of mother function. Furthermore none of the above proposals find the location of strings on sky maps. We recently proposed in~\cite{Ciuca:2017jz} a Bayesian interpretation of cosmic string detection and within that framework we improved on these shortcomings. First of all, our framework would allow different approaches, such as the ones described above, to be unified and studied systematically with machine learning used to search for the optimal combination of methods and choices within each approach. Secondly, we used machine learning to estimate the cosmic string locations in CMB maps,  $\delta_{sky}$, and we derived a connection between these estimates and the probability distribution of the cosmic string tension $G\mu$.  In a machine learning context this is a \textit{supervised classification learning} problem. 
Broadly speaking, we need to build a machine which classifies each pixel in the sky as being on a string or not being on a string. For our case this machine is a convolutional neural network.~\footnote{For a general introductions to machine learning with emphasis on supervised and unsupervised learning, see the textbook by Murphy~\citep{Murphy:1981503} and the one by Hastie et al.~\citep{Hastie:1315326}, and for an introduction to reinforcement learning, see the textbook by Sutton and Barto~\citep{Sutton:td}. For an introduction to neural networks, see the recent book by Goodfellow et. al.~\citep{Goodfellow:2244405}.}  

In this paper we present the details of the convolutional neural network that we developed and used for the analysis presented in~\cite{Ciuca:2017jz}.  We make the code publicly available at \url{https://gitlab.com/oscarhdz/cosmicstringnn_v2} and in the sections that follow, we specify where in the code the procedure being described can be found. In Sections~\ref{supervised} and~\ref{neuralnet} we introduce supervised learning and neural networks, respectively. In Section~\ref{convolutional} we explain the architecture of the particular convolutional neural network we developed. These details are contained in the file {\tt src/model\_def.py} of the code.  The final weights and biases after training are in the file {\tt models/modelv2.pth}. In~\ref{training} we present the methodology used to train our network. The main parts of this procedure are contained in the files {\tt src/train.py}.  A CMB maps passed through our trained network results in a prediction map. But this prediction maps needs to be normalized. This procedure is also discussed in Section~\ref{training} and is contained in file {\tt src/compute\_frequencies.py}.
The calculation of the posterior probability distributions for the string tension $G\mu$ from the normalized prediction maps is done in the file {\tt src/bayesian.py}. 
In Section~\ref{training} we also show that the network performs well with a realistic Nambu-Goto simulation, even though all the development and training was done with a long straight string model.  In Section \ref{conclusions} we present our conclusions and the prospects for future work. 

\section{Supervised Learning}
\label{supervised}
Suppose we have a dataset $D = \{(\mathbfit{x}_i, \mathbfit{y}_i) \, | \, i=1...N\}$, where $\mathbfit{x}_i$ and $\mathbfit{y}_i$ belong to some input space $A$ and some label space $B$, respectively. The goal of supervised learning is to find the mapping $f : A \rightarrow B$ which assigns labels to data points. The space $A$ is frequently $\mathbb{R}^n$, but not always. For example, $A$ could be the space of all graphs with labelled vertices.  Note that we wish to approximate the unknown process which generates the dataset's labels, not fit the dataset itself. A mapping which produces the right labels on the original dataset $D$ but does not produce the right labels on a second similar dataset with previously unseen data produced by the same process is said to be \textit{overfitted}. The standard way to detect overfitting is to split the dataset into a training set $D_{train}$ which we use to find the mapping $f$ and another evaluation set $D_{eval}$ on which we evaluate the performance of the mapping. 

Let $f_\mathbfit{w}: A \rightarrow B$ be a manifold in function space parametrized by  coordinates which we arrange as elements of a vector $\mathbfit{w}$.  
The best choice of $\mathbfit{w}$ is the one which minimizes a problem-dependent error function. In general this function has the following form
\be
\label{error}
	E(\mathbfit{w}) = \sum_{(\mathbfit{x}_i, \mathbfit{y}_i) \in D_{train}} ||\mathbfit{y}_i, f_\mathbfit{w}(\mathbfit{x}_i) || \ \ .
\ee
The symbol $||\ , ||$ is a norm in function space.
For example, if we set the input space $A = B=\mathbb{R} $, and consider the manifold in function space of all linear functions, and set the norm in the last equation to be the squared Euclidean distance, this produces the standard linear regression problem, which
is easily minimized by setting its derivatives to $0$ and solving the resulting system of linear equations for $ f_\mathbfit{w}$. 

The goal of finding the mapping which minimizes~\ref{error} among all functions which map from $A$ to $B$ is of course infeasible since the space is so large. The problem is made tractable by restricting ourselves to a finite dimensional manifold on this function space and finding the best function on the manifold, this can be interpreted as applying an infinitely strong prior on the function space. The field of supervised learning concerns itself with finding the right priors in function space for a given problem. For instance if $A=B=\mathbb{R}^n$, the space of all linear functions from $A$ to $B$ is such a manifold, the prior here is simply zero for all nonlinear functions. Broadly speaking, supervised learning is about designing good functional forms and good error functions for a broad class of problems. In the next section we consider a class of functional forms called neural networks and discuss their properties in some detail.

Casting the problem of predicting string locations on an $m\times m$ pixel CMB map as a supervised learning problem, we make the following assignments: $A = \mathbb{R}^{m \times m}$, $B = \{0,1\}^{m\times m}$. The dataset $D_{train} = \{(\delta^i_{sky}, \xi^i) \, | \, i=1...N\}$ is made up of $N$ simulated sky temperature maps, $\delta^i_{sky}$, with the string map contribution added at different $G\mu$ and the associated string location maps $\xi^i$. Given a pixel $\mathbfit{j}\equiv(j_1,j_2)$, $\xi^i_{j_1,j_2}=1$ or 0, depending on whether a string is located at that pixel or not.  
In this context $f_\mathbfit{w}\equiv \prod_{\mathbfit{j} \in {\rm pixels}} (f_{\mathbfit{w},\, \mathbfit{j}})^{\xi^i_{\mathbfit{j}}}(1-f_{\mathbfit{w},\, \mathbfit{j}})^{1-\xi^i_{\mathbfit{j}}}$ is a convolutional neural network, to be described in Section~\ref{convolutional}, with free parameters labelled by $\mathbfit{w}$. The norm used is the 
Kullback-Leibler divergence (equations~(4.3) and (4.7) in \cite{Ciuca:2017jz}), and in this case the error function is known as the cross entropy. Dropping the unnecessary normalisation factor, we minimize:
\begin{multline}
\label{crossentropy}
E(\mathbfit{w}) =  \sum_{(\delta^i_{sky}, \xi^i) \in D_{train}} ~ \sum_{\mathbfit{j} \in {\rm pixels}} 
- \bigg\{ \xi^i_{\mathbfit{j}}\times \log(f_{\mathbfit{w},\, \mathbfit{j}}(\delta^i_{sky}))
\\
+ (1-\xi^i_{\mathbfit{j}})\times\log (1- f_{\mathbfit{w},\, \mathbfit{j}}(\delta^i_{sky}))
\bigg\}  \, .~~~~~~~~~~~~~~~~~~~~~~~~~~~~
\end{multline}

\section{Neural Networks}
\label{neuralnet}
In general, a neural network is a function from $\mathbb{R}^n$ to $\mathbb{R}^m$ parametrized by some parameter vector $w$. These functions are often made up of alternating linear transformations (or rather affine functions) and element-wise nonlinearities. To obtain the output of the neural network, we first multiply the input $\mathbfit{x} \in \mathbb{R}^n$ by some matrix $\boldsymbol{\mathsf{W}}^1 \in \mathbb{R}^{h_1 \times n}$ whose elements are part of the parameter vector, then we apply an element-wise nonlinear function to the resulting vector $\boldsymbol{\mathsf{W}}^1 \mathbfit{x} \in \mathbb{R}^{h_1}$. The value $h_1$ corresponds to the number of neurons in the first hidden layer. We can repeat this process a number of times to gain some abstraction from the input. This process of alternating linear transformations and element-wise nonlinearities produces "layers" in the functional form. In cases where there are many layers (here many can refer to as little as 3) the network is referred to as ``deep''. The process described above is called deep learning. For a recent review of the field, see \cite{LeCun:2015dt}.

An example which maps from $\mathbb{R}^n \rightarrow \mathbb{R}^m$ using one hidden layer $h_1$ is
\be\label{1layernet}
 F(\mathbfit{x})\;=\; \sigma( \boldsymbol{\mathsf{W}}^2 \sigma(\boldsymbol{\mathsf{W}}^1 \mathbfit{x} + \mathbfit{b}^1) + \mathbfit{b}^2 ) \, ,
\ee
where $\boldsymbol{\mathsf{W}}^1  \in \mathbb{R}^{h_1 \times n}$, $\boldsymbol{\mathsf{W}}^2 \in \mathbb{R}^{m \times h_1}$,  $\mathbfit{b}^1 \in \mathbb{R}^{h_1}$ and  $\mathbfit{b}^2 \in \mathbb{R}^{m}$, $\sigma (x)$ is some nonlinear function which is applied to each element of its vector input. Here the elements of the parameter vector $\mathbfit{w}^i$ are all the elements of $\boldsymbol{\mathsf{W}}^i$ and $\mathbfit{b}^i$, called weights and biases, respectively. If $\sigma$ is a non-constant, bounded, monotonic continuous function such a $\tanh(x)$ or the $\text{sigmoid}(x)\equiv 1/(1+\exp(-x))$, then functions of this class are universal approximators on the unit hypercube \citep{Hornik:ub}. That is, for any piecewise continuous function $G(\mathbfit{x})$ and some $\epsilon >0$, there exists an $h_1$ and some parameters $w$ such that $F(\mathbfit{x})$ approximates $G(\mathbfit{x})$ to a better accuracy than $\epsilon$ (in either the  mean-square or supremum norms). We can generalize the above functional form by allowing for multiple layers, a multi-layered perceptron, (MLP). Let $F({\bf x};\mathbfit{w})$ be of the form above parametrized by the vector of weights and biases $\mathbfit{w}$, then we can define an n-layered neural network as :
\be
\label{mlp}
F_{mlp}({\bf x}) = F_{n}(F_{n-1}(F...F_1({\bf x};\mathbfit{w}^1) ;w...);\mathbfit{w}^{n-1});\mathbfit{w}^n)
\ee
For two layers equation~\ref{mlp} can be written as
$
 F({\bf x})\;=\; \sigma( \boldsymbol{\mathsf{W}}^3 \sigma(\boldsymbol{\mathsf{W}}^2 \sigma(\boldsymbol{\mathsf{W}}^1 {\bf x}+ \mathbfit{b}^1) + \mathbfit{b}^2) + \mathbfit{b}^3 )
$.

Training such a function is the process of parameter fitting to some data using some error characteristic. Let $\boldsymbol{x}_i$ be a set of $N$ inputs and $\boldsymbol{y}_i$ be a set of $N$ answers, where $i$ labels the element in the dataset.  Then we define the error function:
\be
\label{errorfunction}
E(\boldsymbol{\mathsf{W}}) = \sum_{i}^N ||\boldsymbol{y}_i - F_{mlp}({\bf x}_i;\mathbfit{w})||
\ee
If we use the least-square norm or the cross entropy, the error function is differentiable with respect to the parameters $\mathbfit{w}$ and we could in principle exactly optimize by setting the partial derivatives with respect to these parameters to zero and solving the subsequent system of nonlinear equations. 
On the other hand, for a multi-layered function $F$, the error landscape is in general non-convex and has multiple local minima. Though non-trivial, it is nonetheless possible to optimize the error function with various heuristic methods~\citep{Choromanska:ui}, perhaps using a scheme such as gradient descent.

A gradient evaluation of the error function, for a general nonlinear functions of multiple variables, would take the same number of evaluations of the function as the number of weights and biases. 
However, for neural networks there exists an algorithm which allows for gradient computation to be roughly as costly as two passes through the network. 
The algorithm is called backpropagation \citep{Rumelhart:1986er} and is easily derived by applying the chain rule to equation \ref{mlp}. This decrease in the computational cost of evaluating first derivatives is one of the main advantages of neural networks over other function approximators. 

\section{The Convolutional Neural Network's Parametrization}
\label{convolutional}
In this section we describe the convolutional neural network defined in the file {\tt src/model\_def.py} of the our online code. \footnote{\url{https://gitlab.com/oscarhdz/cosmicstringnn_v2}}
The final weights and biases after training are in the file {\tt models/modelv2.pth}.

In~\cite{Ciuca:2017jz} we used a convolutional neural network to produce prediction maps for string locations.  These prediction maps were approximations to the pixel dependent probability $p_{\mathbfit j}$ of there being a string at a pixel $\mathbfit{j}\equiv(j_1,j_2)$ of a sky map, for a given sky map $\delta_{sky}$ and string tension $G\mu$. The prediction map produced by the network depended on series of parameters that we named $\beta$ in~\cite{Ciuca:2017jz} but which we call $\mathbfit{w}$ in this paper's notation. The values of $\mathbfit{w}$ were chosen through the training of the network so that they gave the best approximation to $p_{\mathbfit j}$ as defined through the Kullback-Leibler divergence given in equation 4.7 of reference~\cite{Ciuca:2017jz}. Equation 4.7 from~\cite{Ciuca:2017jz} is the cross entropy we quoted in eq.~\ref{crossentropy} with $f_{\mathbfit{w},\, \mathbfit{j}}(\delta_{sky})$ from eq.~\ref{crossentropy} corresponding to $p_{\beta,\, {\mathbfit j}}(\delta_{sky})$ in eq.~4.7 of ~\cite{Ciuca:2017jz}.  In this section we expand on the nature of this convolutional neural network and explain the choices we made when we designed it. 

The idea behind convolutional neural networks is to exploit the 2 dimensional structure of some data to drastically reduce the number of parameters we need to optimize. Consider using the function \ref{1layernet} for our task. Here the input $\mathbfit{x}$ are the pixel values of the $m\times m$ pixel sky temperature map, and the $\boldsymbol{\mathsf{W}}^1$ matrix will have $m^2 \times h_1$ parameters. The functional form in \ref{1layernet} does not exploit the fact that the input is an image, the network will perform just as well if we randomly permute the pixels in the map. To address this, we impose an infinitely strong prior on the weight matrix $\boldsymbol{\mathsf{W}}^1$ and set most of its entries to $0$. We do this in the following way. First we set $h_1 = m^2$. $\boldsymbol{\mathsf{W}}^1$ now has as  many rows as we have pixels in the CMB maps.  To each row $\mathbfit{i}$ we assign a different pixel in the map. Each column is also already associated with a pixel $\mathbfit{j}$, we impose that the only nonzero entries of $\boldsymbol{\mathsf{W}}^1$ be those where the pixel $\mathbfit{j}$ is in the neighbourhood of pixel $\mathbfit{i}$.  Here the notion of neighbourhood is flexible, but for definiteness say that two pixels are neighbours if they are adjacent.  We impose periodic boundary conditions on our 2-d maps and each pixel then has 8 adjacent pixels. With this definition each row of $\boldsymbol{\mathsf{W}}^1$ will only have $9$ non-zero elements. We also impose that the weight matrix be translation invariant in the sense that  $\boldsymbol{\mathsf{W}}^1_{\mathbfit{i},\mathbfit{j}} = {\boldsymbol{\mathsf{W}}^1}_{\mathbfit{i}+\mathbfit{k},\mathbfit{j}+\mathbfit{k}}$ for every translation by $\mathbfit{k}$. Keep in mind that $\mathbfit{i},\mathbfit{j},\mathbfit{k}$ here are pixels that are represented by pairs of integers.  After these conditions are applied, there remain only $9$ independant parameters in $\boldsymbol{\mathsf{W}}^1$, and $\boldsymbol{\mathsf{W}}^1_{\mathbfit{i},\mathbfit{j}}=0$ for pixels $\mathbfit{j}$ outside of the $3\times 3$ rectangle around $\mathbfit{i}$. 
We see that $\boldsymbol{\mathsf{W}}^1$ is  a convolution, $\boldsymbol{\mathsf{W}}^1[\mathbfit{i}-\mathbfit{j}]\equiv\boldsymbol{\mathsf{W}}^1_{\mathbfit{i},\mathbfit{j}}$ on the sky map, or a filter with kernel size $3\time3$ and stride size 1:
\be\label{scalarconvolution}
(\boldsymbol{\mathsf{W}}^1 \ast\delta_{sky}) [\mathbfit{i}] = 
\sum_{\mathbfit{i'}=1}^{m^2}{\boldsymbol{\mathsf{W}}^1[\mathbfit{i}-\mathbfit{i'}] \, \delta_{sky}[\mathbfit{i'}] }
\equiv
\sum_{\mathbfit{i'}=1}^{m^2} {\boldsymbol{\mathsf{W}}^1}_{\mathbfit{i},\mathbfit{i'}} \, \delta_{sky}[\mathbfit{i'}] 
\ee

We can also have a convolution operation $V$ between vector-valued maps with vector dimensions $A$ to $B$. If each pixel {\bf i} of the map {\bf x} has a vector value of dimension $A$ given by {\bf x}[{\bf i}]$^\alpha$, with $\alpha=1,...A$, then we have the vector-valued map of vector dimension $B$ given by:
\be\label{vectorconvolution}
(V_\beta \ast\mathbfit{x}) [\mathbfit{i}] = \sum_{\alpha=1}^A \sum_{\mathbfit{i'}=1}^{m^2}{V_{\beta \alpha}[\mathbfit{i}-\mathbfit{i'}] \, \mathbfit{x}^{\alpha}[\mathbfit{i'}] }
\ee

We now define 4 convolution operations $\boldsymbol{\mathsf{W}}^a$, $a=1, 2, 3, 4$, between maps with pixels $m\times m= 512\times512$, and with the pixel values having vector-dimensions $A_a, B_a$. We take $A_1=1$ and all the other $A$'s and $B$'s are 32. We also pick a bias vector $\mathbfit{b}^a$ of dimension 32 which we add to every pixel after each convolution. The values of $\boldsymbol{\mathsf{W}}^a_{\alpha\beta}$ and $b^a_{\beta}$ are free parameters that will be chosen by training the network.

The map $(\boldsymbol{\mathsf{W}}^1\ast \delta_{sky})+\mathbfit{b}^1$ will be an $m\times m$ map with vector-dimension $B_1$. On this map we apply $\tanh$ element-wise so that $\tanh(\boldsymbol{\mathsf{W}}^1\ast \delta_{sky}+\mathbfit{b}^1)$ is still an $m\times m$ map with vector-dimension $B_1$. We repeat this process with all 4 convolutions:
\begin{multline}
\label{tanhW4}
(\overrightarrow{\tanh^4( W+b)\ast\delta_{sky}}) \equiv \\
\tanh(\boldsymbol{\mathsf{W}}^4 \ast \tanh(\boldsymbol{\mathsf{W}}^3 \ast  \tanh(\boldsymbol{\mathsf{W}}^2\ast \tanh(\boldsymbol{\mathsf{W}}^1\ast \delta_{sky}+\mathbfit{b}^1)
+\mathbfit{b}^2\, )
+\mathbfit{b}^3\, )
+\mathbfit{b}^4\, )
\nonumber
\end{multline}
Finally we take the dot product of the vector value at each pixel with the vector $\boldsymbol{\mathsf{W}}^5$ to produce a scalar value at each pixel and again add a scalar bias $\mathbfit{b}^5$ to each pixel. 
\bea
{\rm layer\ 1:\ }  \text{~1-dim}&\to\text{32-dim},  {\rm kernel\ size=3, stride=1}
  \nonumber \\
    &\downarrow \tanh \nonumber \\
{\rm layer\ 2:\ }  \text{32-dim} &\to \text{32-dim},  {\rm kernel\ size=3, stride=1}
  \nonumber \\
    &\downarrow \tanh \nonumber \\
{\rm layer\ 3:\ }  \text{32-dim} &\to \text{32-dim},  {\rm kernel\ size=3, stride=1}
  \nonumber \\
    &\downarrow \tanh \nonumber \\
{\rm layer\ 4:\ }  \text{32-dim} &\to \text{32-dim},  {\rm kernel\ size=3, stride=1}
  \nonumber \\
    &\downarrow \tanh \nonumber \\
{\rm layer\ 5:\ }   \text{32-dim} &\to \text{1-dim},  {\rm kernel\ size=1, stride=1} 
\eea
We then apply the $\text{sigmoid}(x)\equiv 1/(1+\exp(-x))$ element-wise on this scalar valued map:
\be\label{fw}
f_{w,\, \mathbfit{j}}(\delta_{sky})\equiv
\text{sigmoid}\bigg(\, 
\boldsymbol{\mathsf{W}}^5\cdot (\overrightarrow{\tanh^4( \mathbfit{W+b})\ast\delta_{sky}})+\mathbfit{b}^5
\, \bigg)
[\mathbfit{j}]\, .
\ee
The weights and biases $\boldsymbol{\mathsf{W}}^1$, $\boldsymbol{\mathsf{W}}^2$, $\boldsymbol{\mathsf{W}}^3$, $\boldsymbol{\mathsf{W}}^4$, $\mathbfit{b}^1$, $\mathbfit{b}^2$, $\mathbfit{b}^3$, $\mathbfit{b}^4$, $\boldsymbol{\mathsf{W}}^5$, $\mathbfit{b}^5$ constitute a total of 
$$\big\{(32\times3^2+32)+(32^2\times3^2+32)\times3+32+1\big\}=28\, 097$$ 
parameters that will be determined when the network is trained. 
These are the parameters we are modifying to optimize the cross entropy defined in eq.~\ref{crossentropy}. The function also has the "hyperparameters" $A_a,B_a$, and the $3\times3$ kernel size which we have fixed in the presentation above. During regression those numbers are kept constant, but we used a small grid search to find which set of hyperparameters would give the best results.
It is possible to understand the form above as the application of $A_a$ different linear filters on the map $\mathbfit x$, obtaining $B_a$ new maps $\mathbfit x'$, 
We then combined them using the local nonlinear function $\tanh$ and sigmoid.  An example of a common operation on maps which has this general form  is the computation of the norm of the gradient: to compute it, we first apply the linear filters corresponding to taking derivatives in $x$ and $y$, we then combine those $2$ derivative maps using a nonlinear function $g(x,y)=\sqrt{(x)^2+(y)^2}$.

We choose this particular functional form for our neural network for several of reasons.  First, by the universal approximation theorem, this form is general enough to correctly approximate any nonlinear filter with support of size $3\times 3$. This generality implies that if there exists a nonlinear filter capable of determining whether a given pixel belongs to a cosmic string by looking at the surrounding $3\times 3$ pixels, then there exists some values of $A$ and $B$ for which~\ref{fw} will be capable of approximating that filter arbitrarily closely. 

Furthermore, noticing that $\tanh(x)$ is approximately linear for small values of $x$, we can see that this form can also approximate linear filters by setting $\boldsymbol{\mathsf{W}}^5$ to high values and all the rest of the $\boldsymbol{\mathsf{W}}^a, \mathbfit{b}^a$ parameters to small values.  We use the $\text{sigmoid}(x)$ function because we wish to interpret the output of $f_{w,\, \mathbfit{j}}(\delta)$ as the probability that a given pixel belongs to a string, probabilities must have values in $(0,1)$. 

Lastly, in image recognition applications, it is common to see functions with multiple convolutional layers (see \cite{Turaga:2010cha} for an example of such a structure). Such functional forms contain nested convolutions and element-wise nonlinearities, as we do. They also often contain so-called ``pooling layers" which make the output invariant under small translations of the input map. We do not use these ``pooling layers'' because we found that they do not work well for our task. String detection from CMB maps requires the extraction of a very small signal from the overwhelming gaussian fluctuations and the smallness of this signal means that functions whose output is invariant under small translations of the input are not good detectors of strings. 

\section{Training and Implementation Details}
\label{training}
As mentioned in Section~\ref{supervised}, the aim of training is to set the free parameters $\mathbfit{w}$ of the network discussed in the last section such that the cross entropy in eq.~\ref{crossentropy} is minimized. The training procedure described below is implemented in the file {\tt src/train.py} of our online code. We used the \textsc{pytorch} environment (\url{https://pytorch.org/}) for machine learning and optimization algorithms. Training the model on a Tesla K80 GPU took 12 h in total.

The minimization is achieved by a gradient descent algorithm which iteratively improves the free parameters in the direction of the negative gradient $\partial E/\partial w_j$.  The parameters $w$ are initially drawn from a distribution with small mean and standard deviation to ensure that the training is better behaved. Starting with small values guarantees that the network is approximately linear at the beginning, so the derivatives will not start out too large or too small. 

Computing $E(\mathbfit{w})$ and $\partial E/\partial w_j$ on all maps in $D_{train}$ at every iteration is prohibitively expensive since $D_{train}$ contains a rather large number of maps. 
This is solved by only sampling some small number of maps and computing the gradient on those. This produces, at every iteration, a noisy estimates of the true gradient.  The more maps we sample at each iteration, the closer the gradient estimate is to the truth. Hence the training proceeds as follows:
\begin{enumerate}\itemsep0.5pt
\item Initiate each element of $\mathbfit{w}$ randomly with a Glorot initialization. Glorot initialization refers to sampling the weights with a standard deviation that depends on the number of filters in the previous layer. The $\mathbfit{w}$'s are taken from a uniform distribution on the interval $[-\sqrt{1/k}/10, \sqrt{1/k}/10]$. Here $k$ is known as the fan-in number and it refers to the square of the kernel size of the current layer, multiplied by the dimension of the vector valued pixel from the previous layer. 
\item Randomly sample $10$ input-answer tuples $(\delta^i_{sky}, \xi^i)$ from $D_{train}$, and compute $f_{\mathbfit{w},\, \mathbfit{j}}(\delta^i_{sky})$ using eq.~\ref{fw} for each.
\item Compute $E(\mathbfit{w})$ over this set of ten $f_{\mathbfit{w},\, \mathbfit{j}}(\delta^i_{sky})$  and $\xi^i$ with eq.~\ref{crossentropy}, and then also compute $\partial E/\partial w_j$. 
\item Set the learning rate or step size $\alpha = 0.001$. Update $w_j$ to $w_j - \alpha \times \partial E/\partial w_j$.
\item Repeat from step 2 until the error function~\ref{crossentropy} no longer decreases. This was achieved after 1000 iterations.  We call $\overline{\mathbfit{w}}$ the value of the parameters $\mathbfit{w}$ obtained after this procedure, and hence the network's output map is $f_{\overline{\mathbfit{w}},\, \mathbfit{j}}(\delta_{sky})$.
\end{enumerate}

In practice it is found that progressively decreasing the $\alpha$ parameter in step 4 is critical for converging to a good solution. There are many competing algorithms for scheduling $\alpha$ updates, for example Momentum, Nesterov, RMSprop, Adagrad, and Adam. See \url{http://ruder.io/optimizing-gradient-descent/}  for a nice review. 
We used the Adam algorithm \citep{2014arXiv1412.6980K} provided in the optimization package of the \textsc{pytorch} environment. This algorithm keeps track of mean and variance of the gradients and uses them to set a parameter-dependent $\alpha$. Adam has two internal free parameters $\beta_1$ and $\beta_2$ that characterize the average timescales of the first and second moments of the gradients respectively (see \url{http://ruder.io/optimizing-gradient-descent/index.html\#adam}). We set to $\beta_1=0.9$ and $\beta_2=0.999$, as recommended by the authors of the programme. These values have been found to make our network adequately convergence. Ideally we would have performed random search over these parameters to obtain better convergence properties, however this would require retraining the network fully a large number of times and is therefore computationally infeasible.

As expected for a probability distribution, the map $f_{\mathbfit{w},\, \mathbfit{j}}(\delta_{sky})$, has values between 0 and 1 for all pixels. This is ensured by the network architecture. However we need to normalize each value to ensure that when the network predicts a probability of a string being in a particular pixel with probability $\phi$, a fraction $\phi$ of them are actually on a string. Thus for a fixed $G\mu$ we do the following:
\begin{enumerate}\itemsep0.5pt
\item Sample $100$ sky maps from $D_{train}$ all with the same $G\mu$ and their associated $\xi$ maps.
\item Obtain $f_{\mathbfit{w},\, \mathbfit{j}}(\delta^i_{sky})$ for each of the sky maps $\delta^i_{sky}$ generated above. Let $f_{i,j}$ be the value of $f_{\mathbfit{w},\, \mathbfit{j}}(\delta^i_{sky})$ at the $i,j$ pixel . 
\item Bin the values $f_{i,j}$ into 1000 bins of size $\Delta x=0.001$. Recall that all values of $f_{i,j}$ are between 0 and 1. For each bin, compute which fraction of the pixels with values between $[x, x + \Delta x]$  contain strings. Let $h_{G\mu}(x)$ be this fraction.
\item Replace the pixel value of $f_{ij}$ by the corresponding $h_{G\mu}(f_{i,j})$ value. In this way we obtain our final prediction map ${\mathbfit p}_{\overline{\mathbfit{w}},\, \mathbfit{j}}(\delta_{sky})
\equiv 
h_{G\mu}(f_{i,j} )$.
\end{enumerate}
We perform the above procedure for 200 values of $G\mu$ between $10^{-11}$ and $10^{-6}$ in equally spaced log intervals: $G\mu=10^{-11+ n \times 5/200}$, $n=0,...,200$. The procedure described above is implemented in the file {\tt src/compute\_frequencies.py} of the code.

To make sure we have not overfitted and to test the performance of our trained network, we evaluated the loss function~\ref{crossentropy} on sky maps produced from gaussian maps and string maps that were not part of the training set. In particular, for the training set, the average cross entropy at the end of training was 0.67. The cross entropy for maps in the evaluation set was indistinguishable from the training set.  

\begin{figure}
\centering
\begin{subfigure}[b]{0.30\textwidth}
\includegraphics[width=1.0\textwidth]{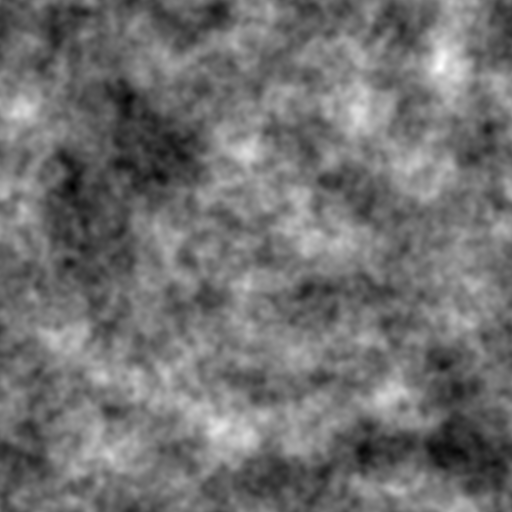}
\caption{The complete sky map, $\delta_{sky}$.}
\label{skymap}
\end{subfigure}
\begin{subfigure}[b]{0.30\textwidth}
\includegraphics[width=\textwidth]{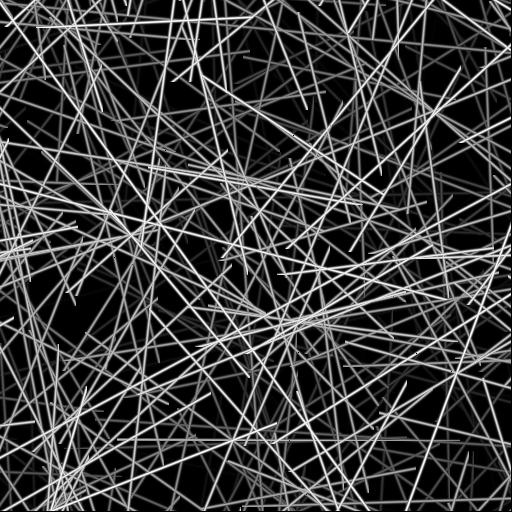}
\caption{The strings' locations, $\xi$.}
\label{strings}
\end{subfigure}
\begin{subfigure}[b]{0.30\textwidth}
\includegraphics[width=\textwidth]{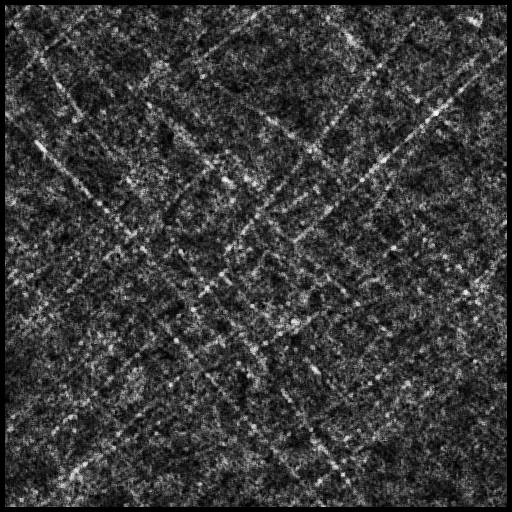}
\caption{ ${\mathbfit p}_{w,\, \mathbfit{j}}(\delta_{sky})$,  $G\mu=10^{-8}$.}
\label{prediction1e-8}
\end{subfigure}
\caption{{\bf Maps from the evaluation set: CMB sky map, cosmic string location map, and a neural network prediction map.}
The maps correspond to $512\times512$ pixels with a resolution of 1 arcmin per pixel. Figure~\ref{skymap} shows a CMB anisotropy temperature map with cosmic strings having a string tension $G\mu=10^{-8}$.
The white and black pixels are $+450\mu$K and $-450\mu$K anisotropies, respectively. CMB maps with and without strings are indistinguishable by eye. 
In \ref{strings} we show the map $\xi$, i.e. the actual placement of long strings in the sky map of figure~\ref{skymap}. The shades of grey of the strings correspond to the relative strength of the CMB temperature discontinuity due to the string. In \ref{prediction1e-8} we show ${\mathbfit p}_{w,\, \mathbfit{j}}(\delta_{sky})$, our neural network's prediction of $\xi$ when the string tension is $G\mu=10^{-8}$. The shades of grey in the prediction maps correspond to the probability of a pixel being on a string, with completely black pixels being 0 probability and completely white pixels being probability 1. }
\label{skymap_fw}
\end{figure}

Using the evaluation set, we produced the map $f_{\mathbfit{w},\, \mathbfit{j}}(\delta_{sky})$ given by eq.~\ref{fw} and apply the procedure above to get our  prediction of the string locations ${\mathbfit p}_{\mathbfit{w},\, \mathbfit{j}}(\delta_{sky})$. Figure~\ref{skymap_fw} shows one such sky map, along with the location of the strings, and the convolutional neural network's prediction ${\mathbfit p}_{\mathbfit{w},\, \mathbfit{j}}(\delta_{sky})$. Figure~\ref{skymap_fw} reproduces figures 1 and 2 from~\cite{Ciuca:2017jz}. Finally, posterior probability distributions for the string tension $G\mu$ are produced from the prediction maps ${\mathbfit p}_{\mathbfit{w},\, \mathbfit{j}}(\delta_{sky})$ through the script defined in {\tt src/bayesian.py}. 

\begin{figure}
\centering
\begin{subfigure}[b]{0.30\textwidth}
\includegraphics[width=1.0\textwidth]{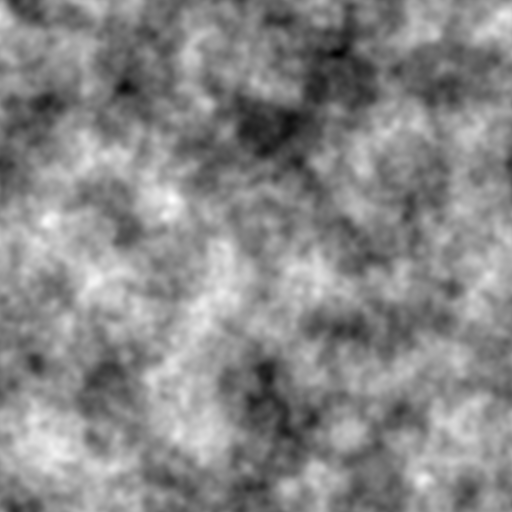}
\caption{The complete sky map, $\delta_{sky}$.}
\label{skyRingeval}
\end{subfigure}
\begin{subfigure}[b]{0.30\textwidth}
\includegraphics[width=\textwidth]{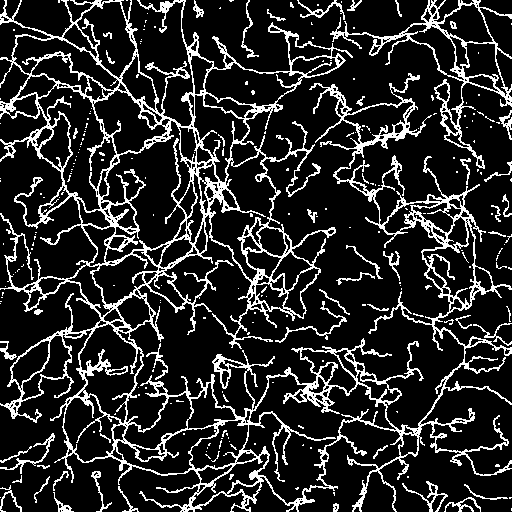}
\caption{The strings' locations, $\xi$.}
\label{stringsRingeval}
\end{subfigure}
\begin{subfigure}[b]{0.30\textwidth}
\includegraphics[width=\textwidth]{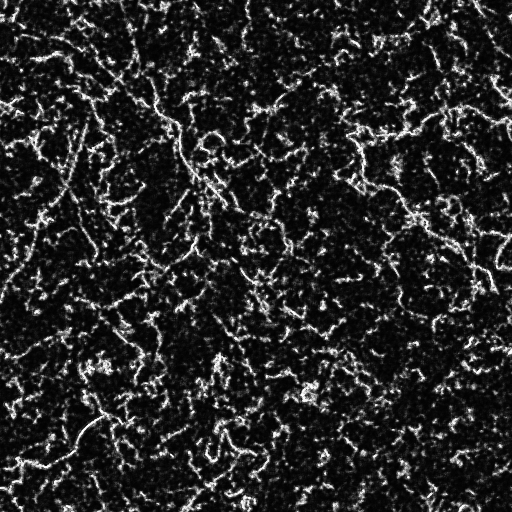}
\caption{ ${\mathbfit p}_{w,\, \mathbfit{j}}(\delta_{sky})$,  $G\mu=5\times10^{-8}$.}
\label{predictionRingeval}
\end{subfigure}
\caption{
Figure~\ref{skyRingeval} shows a CMB anisotropy temperature map with cosmic strings from a realistic Nambu-Goto simulation 
having a string tension $G\mu=5\times10^{-8}$. In \ref{stringsRingeval} we show the actual placement of such strings in the sky map. In \ref{predictionRingeval} we show ${\mathbfit p}_{w,\, \mathbfit{j}}(\delta_{sky})$, our neural network's prediction when analyzing~\ref{skyRingeval}.
}
\end{figure}

Our network was developed and trained within the context of a long string model~\citep{Perivolaropoulos:1993efa}. We described and justified this choice in~\cite{Ciuca:2017jz}. However we have shown that this very same network with no additional training detects strings in realistic Nambu-Goto string simulations without noise.  In figures~\ref{stringsRingeval} and~\ref{predictionRingeval} we show a string location map from a realistic Nambu-Goto simulation~(C.~Ringeval, private communication) similar to those in~\cite{Fraisse:2007nu}, and a prediction map produced by the 4-layer neural network presented here when applied on a sky map containing those strings with string tension $G\mu=5\times10^{-8}$, fig.~\ref{skyRingeval}.  The prediction map was made on a sky map without noise.  These maps are the analogue of those presented in fig.~\ref{skymap_fw}.

Let us explain exactly what we mean when we say we sample from $D_{train}$. Every map in the training set is produced using the following equation:
$
\delta_{sky} = \delta_{gauss} + G_\mu \, \delta_{string}
$.
We numerically produce a dataset of $500$ gaussian maps $\delta^i_{gauss}$ and a dataset of $500$ string temperature maps $\delta^i_{string}$ along with their associated answer maps $\xi^i$.  In the online code these maps are kept in the files {\tt data/cmbedge\_g\_maps.npy}, {\tt data/cmbedge\_s\_maps.npy}, and {\tt data/cmbedge\_a\_maps.npy}, respectively. 
These all have a size of $512\times 512$ pixels with a resolution of 1 arcmin per pixel. We therefore have $500$ unique gaussian maps and $500$ unique string maps.  However, a gaussian map translated with respect to the string map is a distinct gaussian map. Using the translations in 2D, each of the 500 gaussian maps give rise to $512^2$ more gaussian maps. We effectively have $512^2 \times 500$ gaussian maps and thus $512^2 \times 500^2$ distinct combinations of gaussian-string maps.  

There still remains some freedom in the choice of $G\mu$. Every time we want to sample from $D_{train}$ we will generate the $G\mu$ by sampling from a prior distribution $P(G\mu)$. Since Planck quotes a 95\% confidence limit on the Nambu-Goto string tension of $G\mu<1.3\times10^{-7}$, we chose a distribution that decreases exponentially in the string tension with 95\% of its area below $G\mu=9\times10^{-8}$:
\be\label{prior}
P(G\mu) = {\exp({-G\mu / (3\times 10^{-8}}))\over 3\times 10^{-8}} \, .
\ee
Thus the procedure to sample maps from $D_{train}$ is
to first randomly choose a gaussian map, a string map, and a string tension from $\{\delta^i_{gauss}\}, \{\delta^i_{string}\}$ and~\ref{prior}, respectively.  Then we apply a random translation to the gaussian map and finally set $\delta_{sky} = \delta_{gauss} + G_\mu \, \delta_{string}$.  With this $D_{train}$, the network is just going to learn that the large gaussian fields are noise compared to the string map, so the network learns to ignore the gaussian fields and do feature extraction. 

\section{Conclusions}
\label{conclusions}
In~\cite{Ciuca:2017jz} we presented a bayesian interpretation of cosmic string detection and proposed a general machine learning framework which we implemented with a convolution neural network. The network as applied to noiseless simulations of CMB maps with arcmin resolution was able to locate strings and accurately determine the value of the string tension for sky maps having strings with string tension as low as $G\mu=5\times10^{-9}$, which is more than an order of magnitude better than that obtained by the Canny algorithm~\citep{Amsel:2008it, Stewart:2009fr, 2010IJMPD..19..183D}, and significantly better than the $G\mu=3\times10^{-8}$ achieved by the wavelet curvelet approach~\citep{Hergt:2017dr}. It is not straightforward to compare our network's performance to wavelet analysis of ref.~\cite{McEwen:2017cg} which considers full sky maps with noise and at a different resolution. 

Strings locations are visible by eye in our prediction maps for string tensions below $G\mu \sim 10^{-8}$, a feature none of the other approaches achieve.  The prediction maps of string locations produced by our convolutional neural network are unique to our approach as compared to the other methods mentioned above. 
Here we presented the details of the convolutional neural network that produced these prediction maps. We described its structure and explained how it was  trained on simulations of CMB temperature anisotropy maps with and without strings. We also explained in detail how we used it to obtain the prediction maps for the estimated string locations.   In the modern deep learning context our network is relatively simple.  The type of layers and number of layers and parameters used by our network resembles that used for character recognition twenty years ago~\citep{LeCun:wc}.

From a machine learning point of view there is nothing new in the techniques we used. The methods used in this paper are typically applied to image recognition~\citep{Krizhevsky:2012wl} or image segmentation~\citep{Shelhamer:bc} datasets in which the relevant signals are much larger than the signals we are working with. Typical tasks include recognizing a face in an image, whereas the signals we are trying to detect are not even visible by eye.  What is new in our approach is the application to images with such faint features, which in our case are the cosmic strings hidden in the CMB maps. No other machine learning methods have been used with this type of dataset to produce maps resembling our prediction maps. 

The results from this network are very promising and while there are many directions for further research, we wish to mention a couple which are of particular interest to us. 
First of all, we presented our framework within the context of cosmic string detection in CMB temperature anisotropy maps. However it could equally be applied to the detection of cosmic string wakes in 21 cm intensity maps~\citep{Brandenberger:2010hi, Hernandez:2011ima, Hernandez:2012gz, Hernandez:2014cu, daCunha:2016bo}. Applying this framework to 21 cm intensity maps will require modifying the network presented here. 
A second step would be to move away from the long straight string model to realistic Nambu-Goto string simulations similar to those discussed in~\cite{Fraisse:2007nu} and used to place the Planck constraints on the string tension~\citep{PlanckCollaboration:2014il}. 

A final step would be to include noise. Once noise is included, a more sophisticated network will be needed to detect the strings. We believe there is great possibility for the improvement of the network presented here,  both by using better architecture design such as \cite{He:2015tt} and many more layers. Our network used 4 layers and 28,097 parameters, and to train the model it took 12 h on one Tesla K80 GPU.  
The general trend in deep learning research has been to improve network performance with deeper networks.  Modern convolutional networks, such as GoogLeNet~\citep{Szegedy:gt}, which are used for object classification, image segmentation, and evaluating Go board positions, have on the order of $10^2$ layers and $10^7$ free parameters.  Training such networks requires multiple GPUs to be run for days or even weeks. Thus the main limiting factor in our network design will be the access to such computer resources. 

\vspace{-4mm}

\section*{Acknowledgements}
We thank Christophe Ringeval for useful discussions and we thank Fran\c{c}ois Bouchet and Christophe Ringeval for providing the CMB map and the string location map used in figures~\ref{skyRingeval} and~\ref{stringsRingeval}.
We would like to acknowledge the support of the Fonds de recherche du Qu\'ebec -- Nature et technologies (FRQNT) Programme de recherche pour les enseignants de coll\`ege.  Computations were made on the supercomputer Helios from Universit\'e Laval, managed by Calcul Qu\'ebec and Compute Canada. The operation of this supercomputer is funded by the Canada Foundation for Innovation (CFI), the minist\`ere de l'\'Economie, de la science et de l'innovation du Qu\'ebec (MESI) and the Fonds de recherche du Qu\'ebec -- Nature et technologies (FRQNT). 

\bibliographystyle{mnras}
\bibliography{NeuralNet_II_v3published}

\begin{thebibliography}{}
\makeatletter
\relax
\def\mn@urlcharsother{\let\do\@makeother \do\$\do\&\do\#\do\^\do\_\do\%\do\~}
\def\mn@doi{\begingroup\mn@urlcharsother \@ifnextchar [ {\mn@doi@}
  {\mn@doi@[]}}
\def\mn@doi@[#1]#2{\def\@tempa{#1}\ifx\@tempa\@empty \href
  {http://dx.doi.org/#2} {doi:#2}\else \href {http://dx.doi.org/#2} {#1}\fi
  \endgroup}
\def\mn@eprint#1#2{\mn@eprint@#1:#2::\@nil}
\def\mn@eprint@arXiv#1{\href {http://arxiv.org/abs/#1} {{\tt arXiv:#1}}}
\def\mn@eprint@dblp#1{\href {http://dblp.uni-trier.de/rec/bibtex/#1.xml}
  {dblp:#1}}
\def\mn@eprint@#1:#2:#3:#4\@nil{\def\@tempa {#1}\def\@tempb {#2}\def\@tempc
  {#3}\ifx \@tempc \@empty \let \@tempc \@tempb \let \@tempb \@tempa \fi \ifx
  \@tempb \@empty \def\@tempb {arXiv}\fi \@ifundefined
  {mn@eprint@\@tempb}{\@tempb:\@tempc}{\expandafter \expandafter \csname
  mn@eprint@\@tempb\endcsname \expandafter{\@tempc}}}

\bibitem[\protect\citeauthoryear{Amsel, Berger  \& Brandenberger}{Amsel
  et~al.}{2008}]{Amsel:2008it}
Amsel S.,  Berger J.,   Brandenberger R.~H.,  2008, \mn@doi [Journal of
  Cosmology and Astroparticle Physics] {10.1088/1475-7516/2008/04/015}, 2008,
  015

\bibitem[\protect\citeauthoryear{Berndsen, Pogosian  \& Wyman}{Berndsen
  et~al.}{2010}]{Berndsen:2010ku}
Berndsen A.,  Pogosian L.,   Wyman M.,  2010, \mn@doi [Monthly Notices of the
  Royal Astronomical Society] {10.1111/j.1365-2966.2010.16951.x}, 407, 1116

\bibitem[\protect\citeauthoryear{Brandenberger, Danos, Hern{\'a}ndez  \&
  Holder}{Brandenberger et~al.}{2010}]{Brandenberger:2010hi}
Brandenberger R.~H.,  Danos R.~J.,  Hern{\'a}ndez O.~F.,   Holder G.~P.,  2010,
  \mn@doi [Journal of Cosmology and Astroparticle Physics]
  {10.1088/1475-7516/2010/12/028}, 2010, 028

\bibitem[\protect\citeauthoryear{Canny}{Canny}{1986}]{Canny:et}
Canny J.,  1986, \mn@doi [IEEE Transactions on Pattern Analysis and Machine
  Intelligence] {10.1109/TPAMI.1986.4767851}, PAMI-8, 679

\bibitem[\protect\citeauthoryear{Choromanska, Henaff, Mathieu, Ben~Arous  \&
  LeCun}{Choromanska et~al.}{2015}]{Choromanska:ui}
Choromanska A.,  Henaff M.,  Mathieu M.,  Ben~Arous G.,   LeCun Y.,  2015,
  Journal of Machine Learning Research: Workshop and Conference Proceedings,
  38, 192

\bibitem[\protect\citeauthoryear{Ciuca \& Hern{\'a}ndez}{Ciuca \&
  Hern{\'a}ndez}{2017}]{Ciuca:2017jz}
Ciuca R.,  Hern{\'a}ndez O.~F.,  2017, \mn@doi [Journal of Cosmology and
  Astroparticle Physics] {10.1088/1475-7516/2017/08/028}, 2017, 028

\bibitem[\protect\citeauthoryear{Danos \& Brandenberger}{Danos \&
  Brandenberger}{2010}]{2010IJMPD..19..183D}
Danos R.~J.,  Brandenberger R.~H.,  2010, \mn@doi [International Journal of
  Modern Physics D] {10.1142/S0218271810016324}, 19, 183

\bibitem[\protect\citeauthoryear{Fraisse, Ringeval, Spergel  \&
  Bouchet}{Fraisse et~al.}{2008}]{Fraisse:2007nu}
Fraisse A.~A.,  Ringeval C.,  Spergel D.~N.,   Bouchet F.~R.,  2008, \mn@doi
  [Physical Review D] {10.1103/PhysRevD.78.043535}, D78, 043535

\bibitem[\protect\citeauthoryear{Goodfellow, Bengio  \& Courville}{Goodfellow
  et~al.}{2016}]{Goodfellow:2244405}
Goodfellow I.,  Bengio Y.,   Courville A.,  2016, {Deep learning}.
Adaptative computation and machine learning series, The MIT Press, Cambridge,
  MA, \url {http://cds.cern.ch/record/2244405}

\bibitem[\protect\citeauthoryear{Gott}{Gott}{1985}]{Gott:1985eg}
Gott J. R.~I.,  1985, \mn@doi [The Astrophysical Journal] {10.1086/162808},
  288, 422

\bibitem[\protect\citeauthoryear{Hastie, Tibshirani  \& Friedman}{Hastie
  et~al.}{2009}]{Hastie:1315326}
Hastie T.,  Tibshirani R.,   Friedman J.,  2009, {The elements of statistical
  learning: data mining, inference, and prediction; 2nd ed.}.
Springer Series in Statistics, Springer, Dordrecht,
  \mn@doi{10.1007/978-0-387-84858-7}, \url {http://cds.cern.ch/record/1315326}

\bibitem[\protect\citeauthoryear{He, Zhang, Ren  \& Sun}{He
  et~al.}{2015}]{He:2015tt}
He K.,  Zhang X.,  Ren S.,   Sun J.,  2015, arXiv.org, p. arXiv:1512.03385

\bibitem[\protect\citeauthoryear{Hergt, Amara, Brandenberger, Kacprzak  \&
  Refregier}{Hergt et~al.}{2017}]{Hergt:2017dr}
Hergt L.,  Amara A.,  Brandenberger R.~H.,  Kacprzak T.,   Refregier A.,  2017,
  \mn@doi [Journal of Cosmology and Astroparticle Physics]
  {10.1088/1475-7516/2017/06/004}, 2017, 004

\bibitem[\protect\citeauthoryear{Hern{\'a}ndez}{Hern{\'a}ndez}{2014}]{Hernandez:2014cu}
Hern{\'a}ndez O.~F.,  2014, \mn@doi [Physical Review D]
  {10.1103/PhysRevD.90.123504}, 90, 123504

\bibitem[\protect\citeauthoryear{Hern{\'a}ndez \& Brandenberger}{Hern{\'a}ndez
  \& Brandenberger}{2012}]{Hernandez:2012gz}
Hern{\'a}ndez O.~F.,  Brandenberger R.~H.,  2012, \mn@doi [Journal of Cosmology
  and Astroparticle Physics] {10.1088/1475-7516/2012/07/032}, 2012, 032

\bibitem[\protect\citeauthoryear{Hern{\'a}ndez, Wang, Fong  \&
  Brandenberger}{Hern{\'a}ndez et~al.}{2011}]{Hernandez:2011ima}
Hern{\'a}ndez O.~F.,  Wang Y.,  Fong J.,   Brandenberger R.~H.,  2011, \mn@doi
  [Journal of Cosmology and Astroparticle Physics]
  {10.1088/1475-7516/2011/08/014}, 2011, 014

\bibitem[\protect\citeauthoryear{Hornik, Stinchcombe  \& White}{Hornik
  et~al.}{1990}]{Hornik:ub}
Hornik K.,  Stinchcombe M.,   White H.,  1990, Neural Networks, 3, 551

\bibitem[\protect\citeauthoryear{Kaiser \& Stebbins}{Kaiser \&
  Stebbins}{1984}]{Kaiser:1984jg}
Kaiser N.,  Stebbins A.,  1984, \mn@doi [Nature] {10.1038/310391a0}, 310, 391

\bibitem[\protect\citeauthoryear{Kingma \& Ba}{Kingma \&
  Ba}{2014}]{2014arXiv1412.6980K}
Kingma D.~P.,  Ba J.,  2014, arXiv.org, p. arXiv:1412.6980

\bibitem[\protect\citeauthoryear{Krizhevsky, Sutskever  \& Hinton}{Krizhevsky
  et~al.}{2012}]{Krizhevsky:2012wl}
Krizhevsky A.,  Sutskever I.,   Hinton G.~E.,  2012, Advances in Neural
  Information Processing Systems, 25, 1097

\bibitem[\protect\citeauthoryear{LeCun et~al.,}{LeCun et~al.}{1995}]{LeCun:wc}
LeCun Y.,  et~al., 1995, Neural Networks The Statistical Mechanics Perspective,
  pp 261--276

\bibitem[\protect\citeauthoryear{LeCun, Bengio  \& Hinton}{LeCun
  et~al.}{2015}]{LeCun:2015dt}
LeCun Y.,  Bengio Y.,   Hinton G.,  2015, \mn@doi [Nature]
  {10.1038/nature14539}, 521, 436

\bibitem[\protect\citeauthoryear{McEwen, Feeney, Peiris, Wiaux, Ringeval  \&
  Bouchet}{McEwen et~al.}{2017}]{McEwen:2017cg}
McEwen J.~D.,  Feeney S.~M.,  Peiris H.~V.,  Wiaux Y.,  Ringeval C.,   Bouchet
  F.~R.,  2017, \mn@doi [Monthly Notices of the Royal Astronomical Society]
  {10.1093/mnras/stx2268}, 472, 4081

\bibitem[\protect\citeauthoryear{Murphy}{Murphy}{2012}]{Murphy:1981503}
Murphy K.~P.,  2012, {Machine learning: a probabilistic perspective}.
The MIT Press, Cambridge, MA, \url {http://cds.cern.ch/record/1981503}

\bibitem[\protect\citeauthoryear{Perivolaropoulos}{Perivolaropoulos}{1993}]{Perivolaropoulos:1993efa}
Perivolaropoulos L.,  1993, \mn@doi [Physics Letters B]
  {10.1016/0370-2693(93)91825-8}, 298, 305

\bibitem[\protect\citeauthoryear{{Planck Collaboration} et~al.,}{{Planck
  Collaboration} et~al.}{2014}]{PlanckCollaboration:2014il}
{Planck Collaboration} et~al., 2014, \mn@doi [Astronomy and Astrophysics]
  {10.1051/0004-6361/201321621}, 571, A25

\bibitem[\protect\citeauthoryear{Rumelhart, Hinton  \& Williams}{Rumelhart
  et~al.}{1986}]{Rumelhart:1986er}
Rumelhart D.~E.,  Hinton G.~E.,   Williams R.~J.,  1986, \mn@doi [Nature]
  {10.1038/323533a0}, 323, 533

\bibitem[\protect\citeauthoryear{Shelhamer, Long  \& Darrell}{Shelhamer
  et~al.}{2017}]{Shelhamer:bc}
Shelhamer E.,  Long J.,   Darrell T.,  2017, \mn@doi [IEEE Transactions on
  Pattern Analysis and Machine Intelligence] {10.1109/TPAMI.2016.2572683}, 39,
  640

\bibitem[\protect\citeauthoryear{Stewart \& Brandenberger}{Stewart \&
  Brandenberger}{2009}]{Stewart:2009fr}
Stewart A.,  Brandenberger R.~H.,  2009, \mn@doi [Journal of Cosmology and
  Astroparticle Physics] {10.1088/1475-7516/2009/02/009}, 2009, 009

\bibitem[\protect\citeauthoryear{Sutton \& Barto}{Sutton \&
  Barto}{1998}]{Sutton:td}
Sutton R.~S.,  Barto A.~G.,  1998, {Reinforcement Learning: An Introduction}.
MIT Press, \url {http://scholar.google.comjavascript:void(0)}

\bibitem[\protect\citeauthoryear{Szegedy et~al.,}{Szegedy
  et~al.}{2015}]{Szegedy:gt}
Szegedy C.,  et~al., 2015, in 2015 IEEE Conference on Computer Vision and
  Pattern Recognition (CVPR). IEEE, pp~1--9,
  \mn@doi{10.1109/CVPR.2015.7298594}, \url
  {http://ieeexplore.ieee.org/document/7298594/}

\bibitem[\protect\citeauthoryear{Turaga, Murray, Jain, Roth, Helmstaedter,
  Briggman, Denk  \& Seung}{Turaga et~al.}{2010}]{Turaga:2010cha}
Turaga S.~C.,  Murray J.~F.,  Jain V.,  Roth F.,  Helmstaedter M.,  Briggman
  K.,  Denk W.,   Seung H.~S.,  2010, \mn@doi [Neural Computation]
  {10.1162/neco.2009.10-08-881}, 22, 511

\bibitem[\protect\citeauthoryear{Vafaei~Sadr, Movahed, Farhang, Ringeval,
  Bouchet  \& Bouchet}{Vafaei~Sadr et~al.}{2018}]{VafaeiSadr:2018hh}
Vafaei~Sadr A.,  Movahed S. M.~S.,  Farhang M.,  Ringeval C.,  Bouchet F.~R.,
  Bouchet F.~R.,  2018, \mn@doi [Monthly Notices of the Royal Astronomical
  Society] {10.1093/mnras/stx3126}, 475, 1010

\bibitem[\protect\citeauthoryear{da Cunha, Brandenberger  \&
  Hern{\'a}ndez}{da~Cunha et~al.}{2016}]{daCunha:2016bo}
da Cunha D. C.~N.,  Brandenberger R.~H.,   Hern{\'a}ndez O.~F.,  2016, \mn@doi
  [Physical Review D] {10.1103/PhysRevD.93.123501}, 93, 123501

\makeatother
\end{thebibliography}
\bsp	
\label{lastpage}
\end{document}